\documentclass[11pt,envcountsame,%
envcountsect]{llncs}

\usepackage[english]{babel}
\usepackage{amsmath,amssymb,mathdots,amsfonts,enumitem,tikz,wrapfig}
\usepackage[hang,centerlast]{subfigure}
\usepackage[utf8]{inputenc}
\usepackage{tikz}
\usetikzlibrary{automata,calc,positioning}
\usepackage{hyperref}

\tikzstyle{every picture}+=[initial text=]
\tikzstyle{every state}+=[inner sep=1pt,minimum size=15pt]

\newif\ifDraft
\Draftfalse
\usepackage{fixme,manfnt}
\makeatletter
\newcounter{fixcount}
\newcommand{\defineNote}[3][black!65!green]{%
  \expandafter\def\csname @#2\endcsname ##1{\stepcounter{fixcount}\fxwarning{\textcolor{#1}{\textbf{#3}: ##1}}}%
  \expandafter\def\csname @@#2\endcsname ##1{\stepcounter{fixcount}\small\fxwarning[author={\textcolor{red}{#3,}},margin=false,inline=true]{\textcolor{#1}{##1}}}%
  \expandafter\def\csname #2\endcsname{\@ifstar{\csname @@#2\endcsname}{\csname @#2\endcsname}}
}
  \makeatother
\defineNote{BB}{B\'eatrice}
\defineNote{JM}{John}
\defineNote{OK}{Olga}
\defineNote{MS}{Mathieu}

\usepackage{gastex}

\setlist{itemsep=0pt}

\title{Probabilistic Opacity in\\ Refinement-Based Modeling%
\thanks{Partially supported by a grant from Coop\'eration
    France-Qu\'ebec, Service Coop\'e\-ration et Action Culturelle
    2012/26/SCAC (French Government), the NSERC Discovery Individual grant No. 13321 (Government of Canada), the FQRNT Team grant No. 167440 (Quebec's Government) and the CFQCU France-Quebec Cooperative grant No. 167671  (Quebec's Government).}}

\author{~B\'eatrice B\'erard\inst{1,2} and Olga Kouchnarenko\inst{3}
and John Mullins\inst{4}$^,$\thanks{This research has been partially done while this author was visiting the LIP6, Universit\'e Pierre \& Marie Curie.}
and Mathieu Sassolas\inst{5} } 

\institute{Sorbonne Universit\'e, UPMC
Univ Paris 06, UMR 7606, LIP6, Paris, France 
\and CNRS, UMR 7606, F-75005 Paris, France 
\and Universit\'e de Franche-Comt\'e,
FEMTO-ST, CNRS UMR 6174, Inria/NGE, Besan\c{c}on, France 
\and Dept. of Computer \& Software Eng., \'Ecole
Polytechnique de Montr\'eal\\
Montreal (Quebec), Canada, H3C 3A7 \and Universit\'e Paris-Est, LACL,
Cr\'eteil, France}

\newcommand{\prob}{\ensuremath{{\bf P}}}
\newcommand{\dist}{\ensuremath{\mathcal{D}ist}}
\newcommand{\tr}{\ensuremath{\textrm{tr}}}
\newcommand{\Tr}{\ensuremath{\textit{Tr}}}
\newcommand{\FTr}{\ensuremath{\textit{FTr}}}

\newcommand{\lst}{\ensuremath{\textrm{lst}}}
\newcommand{\lang}{\ensuremath{\mathcal{L}}}
\newcommand{\A}{\ensuremath{\mathcal{A}}}

\newcommand{\V}{\ensuremath{\mathcal{V}}}
\newcommand{\Spec}{\ensuremath{\mathcal{S}}}
\newcommand{\obs}{\ensuremath{\mathcal{O}}}
\newcommand{\I}{\ensuremath{\mathcal{I}}}
\newcommand{\Rel}{\ensuremath{\mathcal{R}}}
\newcommand{\sat}{\underline{sat}}
\newcommand{\satgen}{\mathit{sat}}
\newcommand{\FRuns}{\mathit{FRuns}}

\newcommand{\eps}{\ensuremath{\varepsilon}}
\newcommand{\fee}{\ensuremath{\varphi}}

\ifDraft
\fi

\begin{document}

\maketitle

\begin{abstract}
  Given a probabilistic transition system (PTS) $\cal A$ partially
  observed by an attacker, and an $\omega$-regular predicate $\varphi$
  over the traces of $\cal A$, measuring the disclosure of the secret
  $\varphi$ in $\cal A$ means computing the probability that an
  attacker who observes a run of $\cal A$ can ascertain that its trace
  belongs to $\varphi$.
  In the context of refinement, we consider specifications given as
  Interval-valued Discrete Time Markov Chains (IDTMCs), which are
  underspecified Markov chains where probabilities on edges are only
  required to belong to intervals. Scheduling an IDTMC $\Spec$
  produces a concrete implementation as a PTS and we define the worst
  case disclosure of secret $\varphi$ in $\Spec$ as the maximal
  disclosure of $\varphi$ over all PTSs thus produced. We compute this
  value for a subclass of IDTMCs and we prove that refinement can only
  improve the opacity of implementations.
\end{abstract}

\section{Introduction}

\paragraph{Context and motivation.}
When modeling complex systems, a top-down approach based on the
refinement of models allows to gradually specify various system
requirements.  These refinements are designed to preserve some
behavioral properties, like safety, reachability, and liveness under
some conditions.  

Security requirements, which are not behavioral
ones~\cite{clarkson10}, may not fare well under refinement, unless
tailored specially to do so, as in~\cite{alur06}.  Several known
security properties such as noninference or anonymity can be encoded
in the framework of \emph{opacity}~\cite{mazare05,bryans08,alur06}. In
this context, an external observer tries to discover whether a
predicate (given as an $\omega$-regular set) holds by partially
observing the system through a projection of its actions.  A system is
opaque if the attacker fails to discover this information.  In the
possibilistic setting, a violation of opacity captures the existence
of at least one perfect leak.

In probabilistic models like Discrete Time Markov Chains (DTMCs),
naturally random events such as faults, message transmission failure,
actions of an external environment, etc. can be taken into
account. Opacity was extended in the probabilistic
setting~\cite{mscs15,BCS15} to provide a \emph{measure} of the set of
runs \emph{disclosing} information on the truth value of the
predicate.  As a result, opacity increases when the disclosure
--~which is the focus of this paper~-- decreases.

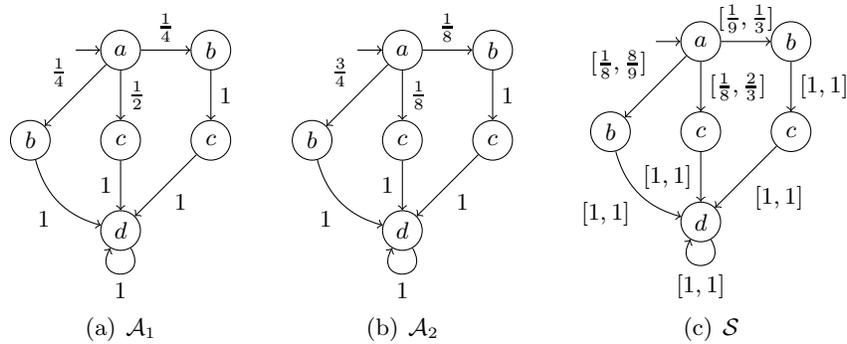
\begin{figure}[h!]
\centering
\tikzset{every loop/.style={min distance=5mm,in=-120,out=-60,looseness=5}}
~\hfill
\subfigure[$\mathcal{A}_1$]{\label{fig:interference1quart}
\begin{tikzpicture}[auto,scale=0.8]
\node[state,initial](s01) at (2.5,5) {$a$};
\node[state] (s11) at (1,3.5) {$b$};
\node[state] (s12) at (2.5,3.5) {$c$};
\node[state] (s13) at (4,5) {$b$};
\node[state] (s14) at (4,3.5) {$c$};
\node[state] (s16) at (2.5,2) {$d$};

\draw[->] (s01) edge[swap] node {$\frac14$} (s11);
\draw[->] (s01) edge node {$\frac12$} (s12);
\draw[->] (s01) edge node {$\frac14$} (s13);

\draw[->] (s11) edge[swap,bend right] node {$1$} (s16);
\draw[->] (s12) edge[swap] node {$1$} (s16);
\draw[->] (s13) edge node {$1$} (s14);
\draw[->] (s14) edge node {$1$} (s16);
\draw[->] (s16) edge[loop below] node {$1$} (s16);
\end{tikzpicture}}
\hfill~
\subfigure[$\mathcal{A}_2$]{\label{fig:interference3quarts}
\begin{tikzpicture}[auto,scale=0.8]
\node[state,initial](s01) at (2.5,5) {$a$};
\node[state] (s11) at (1,3.5) {$b$};
\node[state] (s12) at (2.5,3.5) {$c$};
\node[state] (s13) at (4,5) {$b$};
\node[state] (s14) at (4,3.5) {$c$};
\node[state] (s16) at (2.5,2) {$d$};

\draw[->] (s01) edge[swap] node {$\frac34$} (s11);
\draw[->] (s01) edge node {$\frac18$} (s12);
\draw[->] (s01) edge node {$\frac18$} (s13);

\draw[->] (s11) edge[swap,bend right] node {$1$} (s16);
\draw[->] (s12) edge[swap] node {$1$} (s16);
\draw[->] (s13) edge node {$1$} (s14);
\draw[->] (s14) edge node {$1$} (s16);
\draw[->] (s16) edge[loop below] node {$1$} (s16);
\end{tikzpicture}}
\hfill~
\subfigure[$\Spec$]{\label{fig:interferenceIDTMC}
\begin{tikzpicture}[auto,scale=0.8]
\node[state,initial](s01) at (2.5,5) {$a$};
\node[state] (s11) at (1,3.5) {$b$};
\node[state] (s12) at (2.5,3.5) {$c$};
\node[state] (s13) at (4,5) {$b$};
\node[state] (s14) at (4,3.5) {$c$};
\node[state] (s16) at (2.5,2) {$d$};

\draw[->] (s01) edge[swap] node {$[\frac18,\frac89]$} (s11);
\draw[->] (s01) edge node {$[\frac18,\frac23]$} (s12);
\draw[->] (s01) edge node {$[\frac19,\frac13]$} (s13);

\draw[->] (s11) edge[swap,bend right] node {$[1,1]$} (s16);
\draw[->] (s12) edge[swap] node {$[1,1]$} (s16);
\draw[->] (s13) edge node {$[1,1]$} (s14);
\draw[->] (s14) edge node {$[1,1]$} (s16);
\draw[->] (s16) edge[loop below] node {$[1,1]$} (s16);
\end{tikzpicture}}
\hfill~
%
%
%
%
%
%
%
\caption{Probabilistic systems $\mathcal{A}_1$ or $\mathcal{A}_2$ implementing underspecified system $\Spec$.}
\label{fig:interference}
\end{figure}

Consider for instance the two systems in
\figurename~\ref{fig:interference1quart}-\subref{fig:interference3quarts},
which are DTMCs with the addition of labels on states (indicated
inside). We assume that the occurrence of $b$ must be kept secret and
that all labels except $b$ are observable. In this case, the only runs
\emph{disclosing} the secret are those observed by $ad^{\omega}$,
since every such run betrays the occurrence of $b$.  The probability
of disclosure is $1/4$ in $\A_1$ while it is $3/4$ in $\A_2$, hence
$\A_1$ is more secure than $\A_2$. Our aim is to establish sufficient
conditions on systems like $\mathcal{A}_1$ and $\mathcal{A}_2$, that
can be compared, for one of them to be more secure than the other.


In the process of system modeling, it is common practice to use \emph{underspecified} models as first steps of specification.
A first approach is to consider \emph{sub-stochastic} models where transition probabilities need not sum up to $1$.
In this framework, the notions of satisfaction and simulation were extensively studied in~\cite{baier05}.
The second approach is to introduce non-determinism in the model which makes possible the description of
some choices of the
environment~\cite{larsen-jonsson,segala95,ChatterjeeHenzinger2008,Worrell2013,BCS15}.
These models have also been studied in relation to the refinement
process~\cite{larsen-jonsson}.  For example, both systems of
\figurename~\ref{fig:interference1quart}-\subref{fig:interference3quarts}
could have been derived from a single underspecified system $\Spec$
with the same structure but imprecise probabilities,
like the one in~\figurename~\ref{fig:interferenceIDTMC}.
A particular case of such models is the Interval-valued DTMCs (IDTMCs)
where the transition probabilities are underspecified by giving only
bounds in the form of intervals, as done in
\figurename~\ref{fig:interferenceIDTMC}.
The refinement consists for example in restricting these intervals.
Implementation, on the other hand, reduces the bounds to a single
point, thus providing a Discrete Time Markov Chain.  

Scheduling is an effective way to obtain an implementation: at each
step, the scheduler provides a distribution satisfying the bounds,
producing a (possibly infinite) DTMC on-the-fly.
As such, a scheduler can represent a strategy of an agent inside the
system.  In the case of opacity, this scheduler is adversarial in the
sense that it tries to disclose as much information as possible to the
passive observer.

\paragraph{Contribution.}
In this work, we investigate the effect of refinement of probabilistic
systems on opacity.  Disclosure, as any security measure, is defined
on IDTMCs as the worst case scenario for all its implementations,
although not every implementation is obtained through
scheduling.

In order to evaluate the disclosure of an IDTMC under realistic
assumptions, we study the supremum of the disclosure of all scheduled
implementations: this measures the information obtained by the passive
observer when the system is controlled by the smartest scheduler in
coalition with the observer.  We show how to compute this value for a
subclass of IDTMCs, namely IDTMCs where no transition can be
completely blocked by the scheduler.  The computation is based on
techniques proposed in~\cite{ChatterjeeHenzinger2008}.  Note that the
previous approach in~\cite{BCS15} has already used schedulers to
evaluate disclosure, although in the context of (fully specified) Markov
Decision Processes.

We then show that a refinement of an IDTMC can only improve the
opacity of all implementations obtained by scheduling.  This can be
viewed as an extension of the work in~\cite{alur06} to the
probabilistic setting.
The main difficulty of this result comes from the restriction of the
implementations to those obtained by scheduling.

\paragraph{Organization of the paper.}
In Section~\ref{sec:background} we present the underlying models for
specification and implementation, with the satisfaction relation and
the simulation relation. We define probabilistic disclosure in this
context and show how to compute it (for a restricted case) in
Section~\ref{sec:bg-opacity}.  Finally, we prove monotonicity of
opacity under simulation in Section~\ref{sec:proof}.
Due to lack of space, several proofs are provided in a separate appendix.

\section{Finite automata and probabilistic models for refinement}\label{sec:background}

In this section, we present the models used in this work: finite
automata and the probabilistic models for specifications and
implementations, as well as the satisfaction and simulation relations.

The set of natural numbers is denoted by $\mathbb{N}$. The composition
of relations $\Rel_2$ and $\Rel_1$ is defined by $\Rel_2 \circ \Rel_1
= \{(x,z) \mid \exists y, (x,y) \in \Rel_1 \wedge (y,z) \in
\Rel_2\}$. Given a finite alphabet $\Sigma$, we denote by $\Sigma^*$
(resp. $\Sigma^{\omega}$) the set of finite (resp. infinite) words
over $\Sigma$, with $\Sigma^{\infty} = \Sigma^* \cup \Sigma^{\omega}$
and $\eps$ the empty word.

Given a countable set $Z$, a discrete distribution is a mapping $\mu :
Z \rightarrow [0,1]$ such that $\sum_{z\in Z}\mu(z)=1$. The support of
$\mu$ is $supp(\mu)= \{z \in Z \mid \mu(z)> 0\}$.  The set of all
discrete distributions on $Z$ is denoted by $\dist(Z)$.  When dealing
with a distribution on domain $Z_1 \times Z_2$ (a \emph{joint}
distribution), it is convenient to write $\mu(Y_1,Y_2) = \sum_{y_1 \in
  Y_1, y_2 \in Y_2} \mu(y_1,y_2)$ for $Y_1 \subseteq Z_1$ and $Y_2
\subseteq Z_2$.  Moreover, we abusively write $\mu(y_1,Y_2) =
\mu(\{y_1\},Y_2)$ and $\mu(Y_1,y_2) = \mu(Y_1,\{y_2\})$.

\subsection{Models}
\label{subsec:bg-probproc}

\begin{definition}[Deterministic Parity Automaton]
  A deterministic parity automaton (DPA) is a tuple $\A=(Q,\Sigma,
  \delta, q_0, F)$, where $Q$ is a finite set of states, $\Sigma$ is
  an input alphabet, $\delta:Q\times\Sigma \rightarrow Q$ is a
  transition function, $q_0\in Q$ is an initial state, and $F$ is a
  mapping from $Q$ to a finite set of colors $\{1,\dots,k\}$.
\end{definition}

A run of $\A$ on a word $w=a_1a_2\cdots\in\Sigma^\omega$ is an
infinite sequence $\rho=q_0q_1\cdots \in Q^\omega$ such that for all
$i\geq 0$, $q_{i+1}=\delta(q_i,a_{i+1})$. For such a run $\rho$, we
define $Inf(\rho)$ as the set of states appearing infinitely often in
the sequence. The run is accepting if $min\{F(q)\mid q\in Inf(\rho)\}$
is even. In this case, the corresponding word is accepted by $\A$ and
$\mathcal{L}(\A)$ is the subset of $\Sigma^\omega$ of words accepted
by $\A$.  A subset $K$ of $\Sigma^\omega$ is $\omega$-regular if there
is an automaton $\A$ such that $K=\mathcal{L}(\A)$.

\smallskip For the probabilistic context, we consider two models:
Probabilistic Transition Systems (PTSs) for implementations, and
Interval-valued Discrete Time Markov chains (IDTMCs) for
specifications. 

Probabilistic transition systems are a particular case of the
probabilistic processes of~\cite{larsen-jonsson}, with the restriction
of a countable set of states. They are classical Discrete Time Markov
Chains (DTMCs), with the addition of state labeling.

\begin{definition}[Probabilistic Transition System]
  Let $\Sigma$ be an alphabet. A \emph{probabilistic transition
    system} (PTS) over $\Sigma$ is a $4$-tuple $\A=\langle Q , q_{init},
  \Delta, L \rangle$ where
\begin{itemize}
\item $Q$ is a countable set of states, with $q_{init} \in Q$ the
  initial state,
\item $\Delta: Q \rightarrow \dist(Q)$ is a mapping associating with
  any state $q \in Q$ a distribution $\Delta(q)$ over $Q$, with finite
  support,
\item $L: Q \rightarrow \Sigma$ is the labeling function on states.
\end{itemize}
\end{definition}

A (finite or infinite) run of $\A$ starting from state $q \in Q$ is a
sequence of states $\rho = q_0 q_1 q_2 \ldots$ such that $q_0=q$ and
for each $i$, $0\leq i < |\rho|$, $\Delta(q_i)(q_{i+1}) > 0$. When the
run is finite $\rho = q_0 q_1 \ldots q_n$, we note $q_n = \lst(\rho)$.
The \emph{trace} of $\rho$ is the word $\tr(\rho) = L(q_0)L(q_1)
\ldots \in \Sigma^{\infty}$.  We denote by $Runs_q(\A)$ the set of
infinite runs starting from $q$ and we set $Runs(\A)=
Runs_{q_{init}}(\A)$, and $\Tr(\A) =\{\tr(\rho) \ | \ \rho \in
Runs(\A)\}$, the set of traces of $\A$.  We also define $\FRuns_q(\A)$
the set of finite runs starting from $q$, and similarly $\FRuns(\A) =
\FRuns_{q_{init}}(\A)$ and $\FTr(\A) =\{\tr(\rho) \ | \ \rho \in
\FRuns(\A)\}$, the subset of $\Sigma^*$ of finite traces of $\A$.

Recall~\cite{BillingsleyMeasure95} that a probability measure
$\prob_{\A}$ can be defined on $Runs(\A)$: measurable sets are
generated by \emph{cones}, where the cone $C_\rho$ associated with a
finite run $\rho= q_0q_1 \ldots q_n$ is the subset of infinite runs in
$Runs(\A)$ having $\rho$ as prefix. The probability of $C_\rho$ is
$\prob_{\A}(C_\rho) = \prod_{i=0}^{n-1} \Delta(q_i)(q_{i+1})$.  The
cone of a word $w\in \Sigma^*$ is defined by $C_w = \bigcup_{\rho \in
  \tr^{-1}(w)} C_{\rho}$.  For an $\omega$-regular
language $K$, we denote by $\prob_{\A}(K)$ the probability of $K$ in
$\A$. This probability is well defined (since $\omega$-regular sets
are measurable~\cite{vardi85}) and can be computed given a DPA for
$K$~\cite{courcoubetis95}.

\smallskip The specifications we consider here are given by
Interval-valued Discrete Time Markov chains, introduced
in~\cite{larsen-jonsson} and further investigated
in~\cite{ChatterjeeHenzinger2008,Worrell2013} from a verification
point of view. We denote by $\I$ the set of intervals in $[0,1]$.

\begin{definition}[Interval-valued Discrete Time Markov Chains]
  An Interval-valued Discrete Time Markov Chains (IDTMC) is a
  $4$-tuple $\Spec = (S, s_{init}, T, \lambda)$ where
\begin{itemize}
\item $S$ is a finite set of states, with $s_{init}
  \in S$ the initial state,
\item $T: S \rightarrow (S \rightarrow \I)$ associates with any state
  $s \in S$ a mapping $T(s)$ from $S$ into $\I$,
\item $\lambda: S \rightarrow \Sigma$ is the labeling function.
\end{itemize}
 \end{definition}
 
 By extension, $f \in T(s)$ will denote any function $f: S \rightarrow
 [0,1]$ such that for all $s' \in S$, $f(s') \in T(s)(s')$ and
 $\sum_{s'\in S} f(s') = 1$.

 Several semantic interpretations for IDTMCs  have been considered 
 in~\cite{larsen-jonsson,ChatterjeeHenzinger2008,Worrell2013}. The
 simplest one is Uncertain Markov Chains, which corresponds to first
 choosing all distributions for the states, with probabilities
 belonging to the specified intervals. In this case, the resulting
 model is a PTS, with the same structure as the specification.

 A richer semantics consists in choosing the distribution at each step,
 as in a Markov Decision Process, hence the name Interval Markov
 Decision Process (IMDP).
 Note that an IMDP can be translated into an exponentially larger MDP~\cite{ChatterjeeHenzinger2008}, but the transformed MDP is ill-suited to model refinement.

	 Finally, the most general semantics is directly
 given by the satisfaction relation from~\cite{larsen-jonsson}
 recalled later.

 We first describe the IMDP semantics. A run of $\Spec$ starting from
 a state $s$ is a sequence $s \xrightarrow{\mu_1} s_1
 \xrightarrow{\mu_2} \ldots $ where $s_i \in S$ and each $\mu_i$ is a
 distribution over $S$ such that $\forall s \in S, \mu_i(s) \in
 T(s_{i-1}) (s)$. As before, we denote by $Runs_s (\Spec)$ the set of
 runs starting from $s$, we set $Runs(\Spec) =
 Runs_{s_{init}}(\Spec)$, $\FRuns(\Spec)$ is the set of finite runs of
 $\Spec$ starting from $s_{init}$, and for a run $\rho = s
 \xrightarrow{\mu_1} s_1 \xrightarrow{\mu_2} \ldots s_{n-1}
 \xrightarrow{\mu_n} s_n$ in $\FRuns(\Spec)$ we define
 $\lst(\rho)=s_n$.

 To associate a probability measure with the runs, it is necessary to
 resolve the non determinism by a scheduler that chooses a distribution at
 each step. More precisely:

 \begin{definition}[Scheduler]
   A scheduler $A$ for an IDTMC specification $\Spec= (S, s_{init}, T,
   \lambda)$, is a mapping $A : \FRuns(\Spec)$ $\rightarrow \dist(S)$
   such that for each run $\rho$ with $s=\lst(\rho)$, $A(\rho)(s') \in
   T(s)(s')$.
\end{definition}

We denote by $Sched(\Spec)$ the set of schedulers for $\Spec$. Like
for Markov Decision Processes, scheduling $\Spec$ with $A$ produces a
PTS denoted by $\Spec(A)$ where states are finite runs: $Q \subseteq
\FRuns(\Spec)$, the initial state is the run containing only the
initial state of $\Spec$: $q_{init} = s_{init}$, and for $\rho \in Q$,
$L(\rho) = \lambda(\lst(\rho))$ and $\Delta(\rho)(\rho') =
A(\rho)(s')$ for $\rho' = \rho \xrightarrow{A(\rho)} s'$. We note
$\sat(\Spec)= \{ \Spec(A) \mid A \in Sched(\Spec)\}$.

\smallskip Note that the Uncertain Markov Chains semantics corresponds
to the particular case of memoryless schedulers: there is a mapping $B
: S \rightarrow \dist(S)$ such that $A(\rho) =B(\lst(\rho))$. In this
case the set of states of the PTS $\Spec(A)$ is $Q=S$ with
$\Delta(s)(s') = B(s)(s') \in T(s)(s')$.

\subsection{Satisfaction relation}
\label{subsec:sat}
We now turn to the general satisfaction relation defined
in~\cite{larsen-jonsson}.

 \begin{definition}[Satisfaction relation]
   A PTS $\A=\langle Q , q_{init}, \Delta, L \rangle$ satisfies IDTMC
   $\Spec= (S, s_{init}, T, \lambda)$, written $\A \vDash \Spec$, if
   there exists a relation $\Rel \subseteq Q \times S$ such that
   $q_{init} \Rel s_{init}$ and if $q \Rel s$ then
\begin{enumerate}
\item $L(q) = \lambda(s)$,  
\item there exists a joint distribution $\delta \in \dist(Q\times S)$
 such that 
\begin{enumerate}
\item $\delta(q',S) = \Delta(q)(q')$ for all $q' \in Q$, 
\item $\delta(Q,s') \in T(s)(s')$ for all $s' \in S$, 
\item $q' \Rel s'$ whenever $\delta(q',s')>0$.
\end{enumerate}
\end{enumerate}
\end{definition}

We write $\satgen(\Spec) = \{\A \mid \A \vDash \Spec\}$ for the set of
all PTSs that satisfy specification $\Spec$; these are called \emph{implementations} of $\Spec$.

As a first result, we show that scheduling an IDTMC specification is a particular case of
implementation:

\begin{proposition}
\label{prop-implementation}
Let $\Spec$ be an IDTMC specification. For each scheduler $A$ of
$\Spec$, we have: $\Spec(A) \vDash \Spec$, hence $\sat(\Spec)
\subseteq \satgen(\Spec)$.
\end{proposition}

\begin{proof}
  The relation $\Rel \subseteq Q \times S$ is defined by $\Rel
  =\{(\rho,s)\ | \ \lst(\rho) = s\}$.  We prove that the relation
  $\Rel$ is a satisfaction relation by defining $\delta_{(\rho,s)}$
  over $Q \times S$ as follows:
\begin{equation}
\label{eq}
\delta_{(\rho,s)}(\rho',s') = \left\{ \begin{array}{l@{ }l}
A(\rho)(s') \ &  \ \mbox{if} \ \rho' = \rho \xrightarrow{A(\rho)}  s', \\
0 & \ \mbox{otherwise.}
\end{array}\right.
\end{equation}

The first condition results from the definition of the labeling and
conditions 2 and 3 come from the fact that the joint distribution
$\delta_{(\rho,s)}$ is diagonal in this case.
\begin{itemize}
\item[2] $(a)$ $\delta(\rho',S) = A(\rho)(s') = \Delta(\rho)(\rho')$
  with $\rho' = \rho \xrightarrow{A(\rho)} s'$ for all $\rho' \in Q$.\\
 $(b)$ $\delta(Q,s') = A(\rho)(s') \in T(s)(s')$ with $s =
  \lst(\rho)$ for all $s' \in S$.
\item[3.] $\rho' \Rel s'$ whenever $\delta(\rho',s')>0$ since $s' =
  \lst(\rho')$ by definition of $\Rel$. \qed
\end{itemize}
\end{proof}

Indeed, for any scheduler $A$, $\Spec(A)$ is a kind of
\emph{unfolding} of $\Spec$, which restricts the structure of
$\Spec(A)$: at each step, the scheduler chooses a valid distribution among successor states.
Hence not every implementation can be mapped to a
scheduler.  Said otherwise, not all implementations can be put in
relation with $\Spec$ with a satisfaction relation where the joint
distributions $\delta$ are diagonal. This means that the inclusion
$\sat(\Spec) \subseteq \satgen(\Spec)$ is strict.  

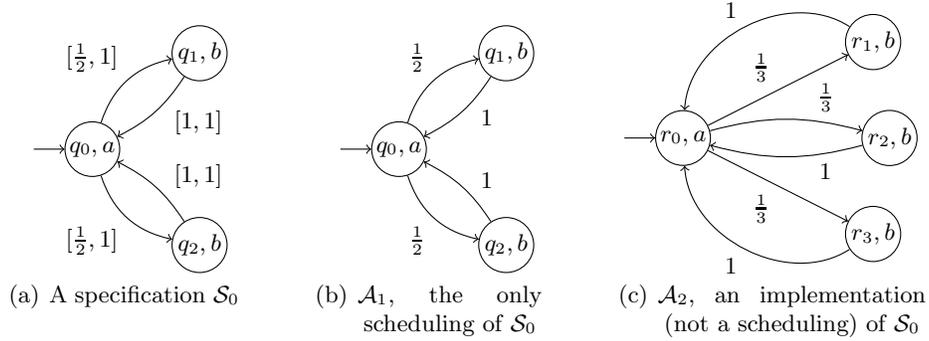
\begin{figure}
\subfigure[A specification $\Spec_0$]{\label{fig:schednotcompletespec}
\begin{tikzpicture}[auto,node distance=0.75cm and 0.9cm]
\node[state,initial] (q0) {$q_0,a$};
\node[state,above right =of q0] (q1) {$q_1,b$};
\node[state,below right =of q0] (q2) {$q_2,b$};

\path[->] (q0) edge[bend left=30] node {$[\frac12,1]$} (q1);
\path[->] (q0) edge[bend right=30,swap] node {$[\frac12,1]$} (q2);
\path[->] (q1) edge[bend left=15,pos=0.33] node {$[1,1]$} (q0);
\path[->] (q2) edge[bend right=15,pos=0.33,swap] node {$[1,1]$} (q0);
\end{tikzpicture}
}
\hfill
\subfigure[$\A_1$, the only scheduling of $\Spec_0$]{\label{fig:schednotcompletesched}
\begin{tikzpicture}[auto,node distance=0.75cm and 0.9cm]
\node[state,initial] (q0) {$q_0,a$};
\node[state,above right =of q0] (q1) {$q_1,b$};
\node[state,below right =of q0] (q2) {$q_2,b$};

\path[->] (q0) edge[bend left=30] node {$\frac12$} (q1);
\path[->] (q0) edge[bend right=30,swap] node {$\frac12$} (q2);
\path[->] (q1) edge[bend left=15,pos=0.33] node {$1$} (q0);
\path[->] (q2) edge[bend right=15,pos=0.33,swap] node {$1$} (q0);
\end{tikzpicture}
}
\hfill
\subfigure[$\A_2$, an implementation (not a scheduling) of $\Spec_0$]{\label{fig:schednotcompleteimplem}
\begin{tikzpicture}[auto,node distance=0.75cm and 2cm]
\useasboundingbox (-0.75,-1.8) rectangle (3.125,1.8);
\node[state,initial] (q0) {$r_0,a$};
\node[state,above right= of q0] (q1) {$r_1,b$};
\node[state,right= of q0] (q2) {$r_2,b$};
\node[state,below right= of q0] (q3) {$r_3,b$};

\path[->] (q0) edge node {$\frac13$} (q1);
\path[->] (q0) edge[bend left=15,pos=0.66] node {$\frac13$} (q2);
\path[->] (q0) edge node[swap] {$\frac13$} (q3);
\path[->] (q2) edge[bend left=15,pos=0.33] node {$1$} (q0);
\path[->] (q1) edge[bend right=60,swap] node {$1$} (q0);
\path[->] (q3) edge[bend left=60] node {$1$} (q0);
\end{tikzpicture}
}

\caption{A specification with an implementation that is not the
  result of scheduling.}
\label{fig:schednotcomplete}
\end{figure}
For example, consider the specification $\Spec_0$ of
\figurename~\ref{fig:schednotcompletespec}.  There is a single possible scheduler for
this specification: the one that picks in $q_0$ probability
$\frac12$ to go to either $q_1$ or $q_2$
($\A_1$ in \figurename~\ref{fig:schednotcompletesched}).  However, the PTS $\A_2$ of \figurename~\ref{fig:schednotcompleteimplem} is also an
implementation of this specification ($\A_2 \models \Spec_0$) where $r_2$ is split between
$q_1$ and $q_2$.  The corresponding matrix is
\[\delta(q_0,r_0) = \bordermatrix{
 & r_0 & r_1 & r_2 & r_3\cr
q_0 & 1 & 0 & 0 & 0 \cr
q_1 & 0 & \frac13 & \frac16 & 0 \cr
q_2 & 0 & 0 & \frac16 & \frac13
}\]

\smallskip Finally, the refinement relation between specifications is
simply defined as inclusion of the satisfaction sets:

 \begin{definition}[Refinement]
   For two IDTMC specifications $\Spec_1$ and $\Spec_2$, $\Spec_1$
   refines $\Spec_2$ if $\satgen({\Spec}_1) \subseteq
   \satgen({\Spec}_2)$.
\end{definition}


\subsection{Simulation relation}\label{subsec:simul}

The notion of simulation relation between probabilistic specifications
was introduced in~\cite{larsen-jonsson}, where it is proved to be a
sufficient condition for refinement: $\Spec_2$ simulates $\Spec_1$
implies that all implementations of $\Spec_1$ are implementations of $\Spec_2$.
This notion is adapted to our setting in Definition~\ref{def-simulation}
below. We then investigate the effect of simulation when
implementations are considered with respect to schedulers only.
Therefore we study the relations between $\sat({\Spec}_1)$ and
$\sat({\Spec}_2)$ whenever $\Spec_2$ simulates $\Spec_1$.

\begin{definition}[Simulation relation]
\label{def-simulation}
For $\Spec_1= (S_1, s_{1,init}, T_1, \lambda_1)$ and $\Spec_2= (S_2,
s_{2,init}, T_2, \lambda_2)$ two IDTMC specifications, $\Spec_2$
simulates $\Spec_1$ if there exists a relation $\Rel \subseteq S_1
\times S_2$ such that $s_{1,init} \Rel s_{2,init}$ and if $s_1 \Rel
s_2$ then:
\begin{enumerate}
\item $\lambda_1(s_1) = \lambda_2(s_2)$, 
\item there exists a function $\delta : S_1 \rightarrow \dist(S_2)$
  such that for all $f \in T_1(s_1)$ and $s'_2 \in S_2$,

\begin{equation} \label{simrel-proba}
\left(\sum_{s'_1 \in S_1}f (s'_1)\cdot \delta (s'_1)(s'_2)\right) \in T_2 (s_2)(s'_2),
\end{equation} 
\item $s'_1 \Rel s'_2$ whenever $\delta(s'_1)(s'_2)>0$.
\end{enumerate}
\end{definition}

\begin{figure}
\begin{center}
\begin{tikzpicture}[auto,node distance=1.125cm and 2cm]
\node[state,initial,label=below:$q_0$,label=left:$\Spec_1:\;\;\;\;$] (q0) {$a$};
\node[state,above right =of q0,label=above:$q_1$] (q1) {$b$};
\node[state,below right =of q0,label=below:$q_2$] (q2) {$b$};

\path[->] (q0) edge[bend left=30,swap] node {$[\frac13,\frac23]$} (q1);
\path[->] (q0) edge[bend right=30] node {$[\frac14,\frac13]$} (q2);
\path[->] (q1) edge[out=300,in=240,loop] node {$[1,1]$} (q0);
\path[->] (q2) edge[out=120,in=60, loop] node {$[1,1]$} (q0);

\node[state,initial,below right=of q1,label=below:$r_0$,label=left:$\Spec_2:\;\;\;\;$] (r0) {$a$};
\node[state,right =of r0,,label=below:$r_1$] (r1) {$b$};

\path[->] (r0) edge node {$[0,1]$} (r1);
\path[->] (r1) edge[out=30,in=330, loop] node {$[1,1]$} (r1);

\path[dashed] (q0) edge[out=75,in=105,distance=3cm] node[below, near end] {$1$} (r0);
\path[dashed] (q1) edge[out=30,in=120,distance=2cm] node[below] {$1$} (r1);
\path[dashed] (q2) edge[bend right=35] node {$1$} (r1);

\end{tikzpicture}
\end{center}

\caption{A simulation of $\Spec_1$ by $\Spec_2$.}
 \label{fig:specsimulation}
\end{figure}
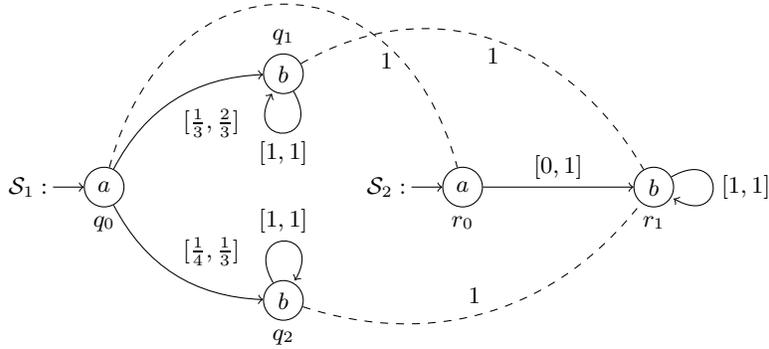

In Figure~\ref{fig:specsimulation}, dashed lines illustrate the
simulation relation $\Rel$ of $\Spec_1$ by $\Spec_2$ labeled with
$\delta(q_i)(r_j)$ since
for Condition~(\ref{simrel-proba}), we may in this case uniformly use
the function:
\[
\delta(q_i)(r_j) = 
\left\{
\begin{array}{ll}
1 & \mbox{if } (q_i, q_j) \in \Rel\\
0 & \mbox{otherwise}
\end{array}
\right.
\]
Note that there is no simulation relation of $\Spec_2$ by
$\Spec_1$. Indeed, let $f \in T_2(r_0)$ defined as $f(r_1) = 1$. The
only way to distribute $f$ over $S_1$ in order to satisfy
Eq.~\ref{simrel-proba} is to distribute $\frac23$ of $f$ to $q_1$ and
$\frac13$ to $q_2$.  Hence, it forces to set $\delta(r_1)(q_1)$ to
$\frac23$ and $\delta(r_1)(q_2)$ to $\frac13$ but this choice for
$\delta$ is not uniform for any $f \in T_2(r_0)$ as it does not
satisfy Eq.~\ref{simrel-proba} for $f(r_1)=\frac12$ for instance.

Let us notice that in the case of PTSs,
Definition~\ref{def-simulation} is still valid but intervals reduce to
points and Equation~\ref{simrel-proba} becomes
\begin{equation} \label{simrel-probaPTSs} \sum_{s'_1 \in S_1}
  \left(\Delta_1(s_1)(s'_1)\cdot \delta (s'_1)(s'_2)\right) = \Delta_2 (s_2)(s'_2).
\end{equation}

\subsection{Simulation \emph{vs} satisfaction}

Remark that not only simulation and satisfaction are defined over different kind of models, they express different requirements.
Indeed, simulation is symmetrical in the models used but asymmetrical in its semantics since one model must contain the behavior of the other.
In contrast, satisfaction concerns different models but the behavior of the PTS must exactly comply to the IDTMC specification: it can neither add behaviors nor introduce new ones.

Nevertheless, these notions coincide when the simulated IDTMC is in fact a PTS:
\begin{proposition}
Let $\A$ be a PTS satisfying an IDTMC specification $\Spec$.
Then $\Spec$ simulates $\A$ when seen as an IDTMC where intervals are reduced to a point.
\end{proposition}

\begin{proof}[Sketch]
The core of the proof relies on building a function $\delta_{sim}$ for the simulation relation, from function $\delta_{sat}$ given by the satisfaction relation.
For any pair of states $q \in Q$ of PTS $\A$ and $s \in S$ of IDTMC $\Spec$, define $\delta_{sim}(q,s) = \frac{\delta_{sat}(q,s)}{\delta_{sat}(q,S)}$.
One can show that this $\delta_{sim}$ satisfies Condition~(\ref{simrel-proba}) of Definition~\ref{def-simulation}.
\end{proof}

Also, as noted before, simulation implies inclusion of the set of implementations:
\begin{proposition}[\cite{larsen-jonsson}]\label{prop:simul-refinement}
$\Spec_2$ simulates $\Spec_1$ implies that $\satgen({\Spec}_1) \subseteq \satgen({\Spec}_2)$.
\end{proposition}

\section{Opacity}\label{sec:bg-opacity}

The original definition of opacity was given in~\cite{bryans08} for
(non probabilistic) transition systems, with respect to some
observation function $\obs$ and some predicate $\fee$ (the secret) on
the runs of the system.

\subsection{Opacity for probabilistic models}
We consider an $\omega$-regular set $\fee \subseteq
\Sigma^{\omega}$ and we say that a run $\rho$ of a PTS $\A$ satisfies $\fee$ if its
trace belongs to $\fee$. We consider also an \emph{observation
  function} $\obs : \Sigma^{\omega}\rightarrow \Sigma_{ob}^{\infty}$,
where $\Sigma_{ob}$ is a subset of $\Sigma$ representing its visible
part. If $\pi_{ob}$ is the standard projection from $\Sigma^{\infty}$
onto $\Sigma_{ob}^{\infty}$ mapping any element $a \in \Sigma
\setminus \Sigma_{ob}$ to $\eps$, the observation function is defined
on a word $w$ by $\obs(w) = \pi_{ob}(w)$. Alternatively,  we could consider a subset
$P_{ob}$ of a set of observable atomic propositions  $P$. Then, by setting $P_u = P
\setminus P_{ob}$, $\Sigma_{ob} = 2^{P_{ob}}$ and $\Sigma_{u} =
2^{P_{u}}$, $\Sigma$ could be viewed as the product $\Sigma_{ob} \times
\Sigma_u$. The observation function $\obs$ would be then defined on a word
$w$ by $\obs(w) = \pi_1(w)$, where $\pi_1$ is the projection from
$(\Sigma_{ob} \times \Sigma_u)^{\infty}$ onto $\Sigma_{ob}^{\infty}$
mapping each element on its first component.

The observation class of a word $w$ is $[w]_{\obs}=
\obs^{-1}(\obs(w))$: for the observer, each word in this class is undistinguishable
from $w$.  Then $\fee$ is \emph{opaque} with respect to $\A$ and
$\obs$ if each time a word satisfies $\fee$, another word in the same
observation class does not.
We denote by $\V(\A,\obs,\fee)$ the set of words violating this condition:
\[\V(\A,\obs,\fee)=(\Tr(\A)\cap\varphi)\setminus(\obs^{-1}(\obs(\Tr(\A)\setminus\varphi))).\]

\begin{definition}[Opacity]
  Let $\A$ be a PTS, with observation function $\obs$. A predicate
  $\fee$ is \emph{opaque} in $\mathcal{A}$ for $\obs$ if
  $\V(\A,\obs,\fee)=\emptyset$.
 \end{definition}

 Equivalently, $\fee$ is opaque if for any $w$ satisfying $\fee$,
 $[w]_{\obs} \nsubseteq \fee$.  Variants of opacity have been
 defined, with other observation functions and predicates, or by
 requiring symmetry: the predicate $\varphi$ is \emph{symmetrically
   opaque} in $\mathcal{A}$ for $\obs$ if both $\varphi$ and
 $\Sigma^\omega\setminus\varphi$ are opaque.

 The notion of probabilistic opacity~\cite{qest10} extends this
 boolean property by defining a measure of the set of runs violating
 opacity:

 \begin{definition}[Probabilistic Disclosure]
   Let $\A$ be a PTS, with observation function $\obs$ and
   $\omega$-regular predicate $\fee$.  The \emph{probabilistic
     disclosure} of $\fee$ in $\mathcal{A}$ for $\obs$ is
   $Disc(\A,\obs,\fee)=\prob_{\A}(\V(\A,\obs,\fee))$.
 \end{definition}

 For instance, recall systems $\A_1$ and $\A_2$ of
 \figurename~\ref{fig:interference}.  The secret predicate in this
 case is the set of runs $\fee_b=ab\Sigma^\omega$ and the observation
 function is the projection $\pi_{\{a,c,d\}}$ onto $\{a,c,d\}^\omega$.
 This predicate is not opaque since the run $abd^\omega$ discloses the
 occurrence of $b$.  This is measured by the disclosure:
 $Disc(\A_1,\pi_{a,c,d},\fee_b)=\prob_{\A_1}(abd^\omega)=\frac14$ and
 $Disc(\A_2,\pi_{a,c,d},\fee_b)=\prob_{\A_2}(abd^\omega)=\frac34$.

 Note that dealing with symmetric opacity would simply require to add
 both measures for $\fee$ and $\Sigma^\omega\setminus\varphi$.  Also
 remark that disclosure only measures probabilities of the observer
 being \emph{sure} that the run is in the secret.
 For example, one can model anonymity of an agent $\alpha$ initiating some protocol by defining $\fee_\alpha$ as the set of all runs initiated by $\alpha$.
 Anonymity of $\alpha$ is then equivalent to opacity of $\fee_\alpha$.
 In the case where anonymity is not guaranteed, disclosure provides a measure of the threat.
 In the case where anonymity holds, this measure will be $0$ and does not give any insight on the ``strength'' of anonymity.
 Other notions measuring this strength were proposed in~\cite{chaum88,bhargava05} and quantitative opacity for partial disclosure of the secret have also been defined in~\cite{mscs15}, although they are not linear hence do not fare well under standard optimization techniques.



\smallskip In order to compare systems with respect to opacity, we
consider two PTSs $\A_1$ and $\A_2$ over the same alphabet $\Sigma$,
with a given observation function $\obs$.  We say that $\A_1$ is more
opaque than $\A_2$ if $Disc(\A_1,\obs,\fee) \leq
Disc(\A_2,\obs,\fee)$.

We lift the definition of probabilistic disclosure from PTSs to IDTMC
specifications as follows:
\[Disc(\Spec,\obs,\fee) = \sup_{\A \in \satgen(\Spec)}
Disc(\A,\obs,\fee)\] From the results in~\cite{larsen-jonsson}
mentioned in Proposition~\ref{prop:simul-refinement}, it is easy to
see that if $\Spec_2$ simulates $\Spec_1$, then
$Disc(\Spec_1,\obs,\fee) \leq Disc(\Spec_2,\obs,\fee)$.

To obtain a similar result for the notion of satisfaction restricted
to schedulers, we define the restriction of the disclosure to 
scheduled implementations: 
\[\underline{Disc}(\Spec,\obs,\fee) = \sup_{\A \in \sat(\Spec)}
Disc(\A,\obs,\fee) = \sup_{ A \in Sched(\Spec)}
Disc(\Spec(A),\obs,\fee)\]

Note that this notion differs from the similar one in~\cite{BCS15} for
Markov Decision Processes.  The notion presented here is finer since
the set of runs measured by the disclosure depends on the scheduled
implementation.  In~\cite{BCS15}, the set of runs of the disclosure is
defined on the (unscheduled) MDP, and its probability is optimized
afterwards.  This would not be consistent in IDTMCs, since two
scheduled implementations can have different sets of edges with
non-null probability, as explained below.

\subsection{Computing the probabilistic disclosure of a specification}

When the interval of an edge of an IDTMC is non-punctual and closed on
the left by $0$, then this edge may be present or not in an
implementation. Otherwise said, in the case of schedulers, this action
can be completely blocked.  In keeping with the terminology
of~\cite{larsen-jonsson}, we call these edges \emph{modal} edges, and
IDTMCs that contain such edges are also called modal IDTMCs.

\begin{definition}[Modal edge]
An edge $T(s)(s')$ in IDTMC $\Spec$ is \emph{modal} if there exists a scheduler $A$ such that in $\Spec(A)$, for any run $\rho$ with $\lst(\rho)=s$, $\Delta(\rho)(\rho\xrightarrow{A(\rho)} s')=0$.
\end{definition}

From a modeling point of view, modal edges add a lot of flexibility for refinement.
This however means that the range of potential implementation is larger and so it will be harder to obtain meaningful properties. Therefore such edges are desirable in an early modeling phase but less so in the latest refinements.
In the context of opacity, removing an edge drastically changes the
disclosure, since it can remove ambiguities.

\begin{figure}[b!]
\tikzset{every loop/.style={min distance=5mm,in=45,out=105,looseness=5}}
\centering
\subfigure[A modal IDTMC $\Spec_{\textrm{m}}$.]{\label{fig:modalImdp}%
\begin{tikzpicture}[auto]
\node[state,initial] (init) at (0,0) {$a$};
\node[state] (sec) at (1.75,0) {$c$};
\node[state] (fin) at (3.5,0) {$b$};

\path[->] (init) edge node {$[0;1]$} (sec)
                 edge[bend left=50] node {$[0;1]$} (fin);
\path[->] (sec) edge node {$[1;1]$} (fin);
\path[->] (fin) edge[loop above] node {$[1;1]$} (fin);
\end{tikzpicture}
}
~~~~~~
\subfigure[A non-modal IDTMC $\Spec_{\textrm{nm}}$.]{\label{fig:noModalImdp}%
\begin{tikzpicture}[auto]
\node[state,initial] (init) at (0,0) {$a$};
\node[state] (sec) at (1.75,0) {$c$};
\node[state] (fin) at (3.5,0) {$b$};

\path[->] (init) edge node {$]0;1]$} (sec)
                 edge[bend left=50] node {$]0;1]$} (fin);
\path[->] (sec) edge node {$[1;1]$} (fin);
\path[->] (fin) edge[loop above] node {$[1;1]$} (fin);
\end{tikzpicture}
}

\subfigure[A disclosing implementation of $\Spec_{\textrm{m}}$ (but not of $\Spec_{\textrm{nm}}$).]{\label{fig:implModal}%
\begin{tikzpicture}[auto]
\node[state,initial] (init) at (0,0) {$a$};
\node[state] (sec) at (1.75,0) {$c$};
\node[state] (fin) at (3.5,0) {$b$};

\path[->] (init) edge node {$1$} (sec);
\path[->] (sec) edge node {$1$} (fin);
\path[->] (fin) edge[loop above] node {$1$} (fin);
\end{tikzpicture}
}
~~~~~~
\subfigure[A non-disclosing implementation of $\Spec_{\textrm{nm}}$ (and $\Spec_{\textrm{m}}$), $\varepsilon>0$.]{\label{fig:implNoModal}%
\begin{tikzpicture}[auto]
\node[state,initial] (init) at (0,0) {$a$};
\node[state] (sec) at (1.75,0) {$c$};
\node[state] (fin) at (3.5,0) {$b$};

\path[->] (init) edge node {$1-\varepsilon$} (sec)
                 edge[bend left=50] node {$\varepsilon$} (fin);
\path[->] (sec) edge node {$1$} (fin);
\path[->] (fin) edge[loop above] node {$1$} (fin);
\end{tikzpicture}
}
\caption{The influence of modal transitions on disclosure.}
\label{fig:modality}
\end{figure}
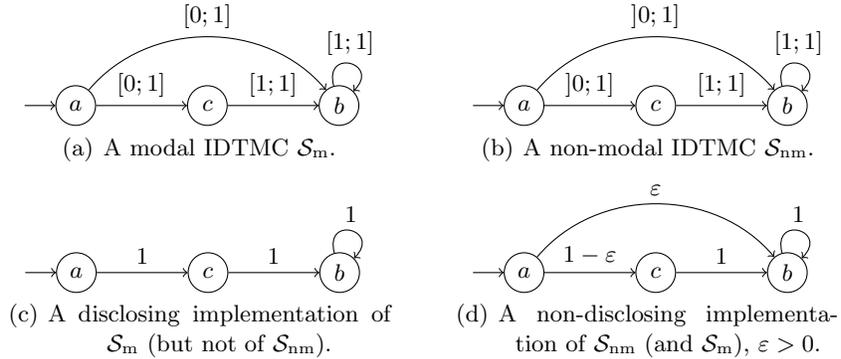

We illustrate the importance of modal edges when computing the
disclosure of a specification.  For example, consider the modal IDTMC
$\Spec_{\textrm{m}}$ of \figurename~\ref{fig:modalImdp}, with an
observation function that observes $a$ and $b$ and a secret being the
presence of $c$.  An implementation of $\Spec_{\textrm{m}}$ that
blocks the direct edge from $a$ to $b$
(\figurename~\ref{fig:implModal}) has a disclosure of $1$, since the
secret is guaranteed to be part of the only possible run.  On the
other hand, in the non-modal version of the IDTMC
(\figurename~\ref{fig:noModalImdp}), such implementations are banned
and only implementations that retain a small probability to avoid $c$
are allowed.  In these implementations, the disclosure is $0$, since
every run is observed as $ab^\omega$ and it is possible that $c$ did
not occur.

In the case of non-modal IDTMCs, disclosure can be computed.
The proof relies on a translation from IMDP to MDP and synchronization with DPA; it can be found in Appendix~\ref{app:computDisc}.
\begin{theorem}\label{thm:computNoModal}
  Computing the value of disclosure for an IDTMC specification $\Spec$
  without modal edges can be done in 2EXPTIME.
\end{theorem}

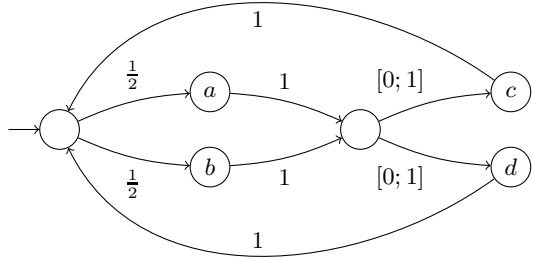
\begin{figure}[h]
\centering
\begin{tikzpicture}[auto]

\node[initial,state] (orig) at (0,0) {};
\node[state] (a) at (2,0.5) {$a$};
\node[state] (b) at (2,-0.5) {$b$};
\node[state] (post) at (4,0) {};
\node[state] (c) at (6,0.5) {$c$};
\node[state] (d) at (6,-0.5) {$d$};

\path[->] (orig) edge[bend left=10,pos=0.66] node {$\frac12$} (a);
\path[->] (orig) edge[bend right=10,pos=0.66,swap] node {$\frac12$} (b);
\path[->] (a) edge[bend left=10,pos=0.33] node {$1$} (post);
\path[->] (b) edge[bend right=10,pos=0.33,swap] node {$1$} (post);
\path[->] (post) edge[bend left=10] node {$[0;1]$} (c);
\path[->] (post) edge[bend right=10,swap] node {$[0;1]$} (d);
\path[->] (c) edge[bend right=40,in=-120] node {$1$} (orig);
\path[->] (d) edge[bend left=40,in=120,swap] node {$1$} (orig);
\end{tikzpicture}
\caption{IDTMC where the choice on modal edge requires history.}
\label{fig:histmodaledge}
\end{figure}

\paragraph{Remarks on modal edges.}
When a scheduler is faced with the choice to include or exclude a modal edge, it can produce several versions of PTSs, say $\A_1$ and $\A_2$, with $\Tr(\A_1)\neq\Tr(\A_2)$, hence $\V(\A_1,\obs,\fee)\neq\V(\A_2,\obs,\fee)$.
In addition, these choices may be history dependent, as in the example of \figurename~\ref{fig:histmodaledge}, with $\varphi=a\Sigma^\omega$ and only letters $c$ and $d$ being observed.
Intuitively, a way for the scheduler to always disclose the presence of an initial $a$ is to always follow an $a$ by the same letter, say a $c$.
However, this choice must be made after the first letter has been seen.
Moreover, leaving the possibility of a run $ad\cdots$ to occur means that run $ac\cdots$ does not disclose $\varphi$.
As a result, the scheduler should also take into account $\varphi$ and the observation function before committing to a choice with respect to modal edges.
So far, the general case of modal IDTMCs remains open.

\section{Monotonicity of opacity under simulation}\label{sec:proof} 

This section is devoted to the proof of the following result: 

\begin{theorem}\label{th:simulation-opacity}
  Let $\Spec_1$ and $\Spec_2$ be IDTMC specifications such that
  $\Spec_2$ simulates $\Spec_1$. Then
  $\underline{Disc}(\Spec_1,\obs,\fee) \leq
  \underline{Disc}(\Spec_2,\obs,\fee)$.
\end{theorem}

Since scheduling is a restrictive way to derive implementations from a
specification, it is not the case in general that $\sat(\Spec_1)
\subseteq \sat(\Spec_2)$: although any scheduling $\Spec_1(A_1)$ of
$\Spec_1$ with $A_1$ is an implementation of $\Spec_2$ (by
Propositions~\ref{prop-implementation}
and~\ref{prop:simul-refinement}), this implementation may not be a
scheduling (as illustrated by the example of
\figurename~\ref{fig:schednotcomplete}).

\begin{wrapfigure}[12]{r}{6.cm}
\centering
\begin{tikzpicture}[auto, node distance=2cm and 3cm]
\useasboundingbox (-0.875,-0.25) rectangle (4.5,-2.75);
\node (s2) at (0,0) {$\Spec_2$};
\node[right = of s2] (s1) {$\Spec_1$};
\node[below = of s1] (impl1) {$\Spec_1(A_1)$};
\node[below = of s2] (impl2) {$\Spec_2(A_2)$};

\path[->] (s2) edge node {$\Rel$} (s1);
\path[->] (s1) edge node {$sat_1$} (impl1);
\path[->,dotted] (s2) edge node {$sat_1\circ\Rel$} (impl1);
\path[->,dashed] (impl2) edge node {$\Rel'$} (impl1);
\path[->,dashed] (s2) edge[swap] node {$sat_2$} (impl2);

\end{tikzpicture}
\caption[The result of Theorem~\ref{thm-refinement}.]{The result of Theorem~\ref{thm-refinement}. Relation $sat_1\circ\Rel$ always exists but might not be a scheduling.}
\label{fig:diagPreuve}
\end{wrapfigure}
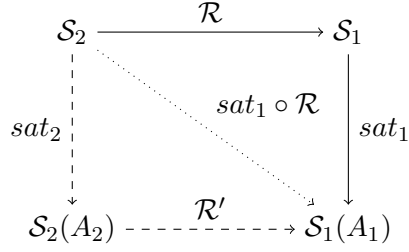

Instead, the proof builds a scheduler $A_2$ for $\Spec_2$ that
produces an implementation $\Spec_2(A_2)$ that \emph{simulates}
$\Spec_1(A_1)$ (Theorem~\ref{thm-refinement}, illustrated in
\figurename~\ref{fig:diagPreuve}).  Then, this simulation is shown to
ensure that the probabilities of (cones of) finite words coincide
(Propositions~\ref{prop-TraceEquiv} and~\ref{prop-TraceEquivValue}).
Disclosure being a measurable event, coincidence of probabilities on
cones ensures coincidence of probabilities for the disclosure.
\medskip

\noindent\textbf{Notations.}
Given two specifications $\Spec_1$ and $\Spec_2$ such that $\Spec_2$
simulates $\Spec_1$ through relation $\Rel$, we define the relation
$\sim$ on $\FRuns(\Spec_1) \times \FRuns(\Spec_2)$ by: $\rho_1 \sim
\rho_2$ if $|\rho_1| = |\rho_2|$ and at any intermediate step $i$, the
corresponding states satisfy $s_{1,i} \Rel s_{2,i}$.

Let $A_1$ and $A_2$ be two schedulers of $\Spec_1$ and $\Spec_2$,
respectively.  We set $\A_1 = \Spec_1(A_1)$ and $\A_2 = \Spec_2(A_2)$,
with respective sets of states $Q_1$ and $Q_2$.  For $\rho_2 \in Q_2$,
we set $sim(\rho_2) = \{ \rho_1 \in Q_1 \mid \rho_1 \sim \rho_2\}$. We
now define a measure $\mu_{\rho_2}$ over $sim(\rho_2)$ by
$\mu_{\rho_2}(\rho_1)=
\frac{\prob_{\A_1}(\rho_1)}{\prob_{\A_1}(sim(\rho_2))}$ (where the
probability of finite run $\rho$ is abusively written instead of the
probability of its cone $C_\rho$).

\medskip

We first show how to build a scheduler for $\Spec_2$ that simulates
the scheduling of $\Spec_1$.  The proof of the theorem below is given
in Appendix~\ref{app:refinement}.
\begin{theorem}
\label{thm-refinement}
Let $\Spec_1$ and $\Spec_2$ be IDTMC specifications such that
$\Spec_2$ simulates $\Spec_1$. Then for any $A_1 \in Sched(\Spec_1)$
there exists $A_2 \in Sched(\Spec_2)$ such that $\Spec_2(A_2)$
simulates $\Spec_1(A_1)$.
\end{theorem}

%
%

\medskip

Now we show that simulation between two PTSs is sufficient to compare
their disclosure.  Namely, we show that the probabilities of cones of
words are equal in both systems.
Note that although this property is well known to hold for paths, it needs to be lifted to words in order to compare disclosure.

We start by considering the sets of traces globally; although it is folklore that simulation implies trace inclusion, we provide a proof for completeness sake.

\begin{proposition}
\label{prop-TraceEquiv}
Let $\A_1$ and $\A_2$ be  PTSs such that $\A_2$ simulates $\A_1$.
Then $Tr(\A_1) = Tr(\A_2)$.
\end{proposition}

\begin{proof}
  We prove the proposition by induction on a strengthened statement.
  Namely, we claim that for every finite run in $\A_1$ there exists a
  similar run in $\A_2$.  Since an infinite run is the limit of the
  sequence of its finite prefixes, this claim is sufficient to prove
  the proposition.  Assume by induction that the proposition holds for
  every word of length $n$.  Let $w \in \FTr(\A_1)$ of length $n+1$.
  We write $w = w_0 a$ for some $a \in \Sigma$.  Consider a run of
  $\A_1$ that produces $w$.  It is of the form $\rho_0 s_1'$ where
  $\lambda(s_1') = a$; let $s_1 = \lst(\rho_0)$.  Let $\rho_0'$ be a
  run in $\A_2$, similar to $\rho_0$, and $s_2 = \lst(\rho_0')$.  By
  definition of simulation, there exists a function $\delta$ such that
  for any state $s_2'$ of $\A_2$,
\begin{equation*}
  \Delta_2 (s_2)(s'_2) = \sum_{\sigma_1 \in S_1} \Delta_1(s_1)(\sigma_1)\cdot \delta(\sigma_1)(s'_2).
\end{equation*}
Moreover, whenever $\delta(\sigma_1)(s'_2)>0$, $\lambda(s_1') =
\lambda(s_2')$.  Since $\delta(s_1')$ is a distribution over $S_2$,
$\delta(s'_1)(s'_2)>0$ for at least one state $s_2'$.  Hence $\rho_0'
s_2'$ is similar to $\rho$, which shows in particular that $w \in
\FTr(\A_2)$.  \qed
\end{proof}

We additionally show that probabilities coincide:
\begin{proposition}
\label{prop-TraceEquivValue}
Let $\A_1$ and $\A_2$ be  PTSs such that $\A_2$ simulates $\A_1$.
Then for all $w\in \Sigma^*$, $\prob_{\A_1}(C_w) = \prob_{\A_2}(C_w)$. 
\end{proposition}
Since a given word may be produced by several paths, their probabilities should be considered altogether.
Hence the proof of the above proposition is not immediate; it is quite technical and can be found in
Appendix~\ref{app:proof:prop-TraceEquivValue}.

\medskip

Existing properties about simulation for PTSs can be retrieved as consequences of the above result.
They were for example obtained as a particular case of sub-stochastic simulation in~\cite{baier05}.
Although not necessary to prove the main theorem, these results illustrate how constraining simulation between PTSs is.

Recall that a probabilistic bisimulation~\cite{larsen-jonsson} is a
bisimulation that preserves transition probabilities, \emph{i.e.}  a
bisimulation relation $\Rel$ on states such that for any equivalence
class $R$ of $\Rel$, and any two related states $s\Rel s'$,
$\Delta(s)(R)=\Delta(s')(R)$.

\begin{corollary}[\cite{baier05}]\label{cor:prob-bisimulation}
Let $\A_1$ and $\A_2$ be PTSs such that $\A_2$ simulates $\A_1$.
Then there exists a probabilistic bisimulation over the union of both PTSs.
\end{corollary}


\begin{corollary}[\cite{baier05}]
Let $\A_1$ and $\A_2$ be  PTSs such that $\A_2$ simulates $\A_1$.
Then $\A_1$ also simulates $\A_2$.
\end{corollary}



We are now ready to prove Theorem~\ref{th:simulation-opacity}.
\begin{proof}
  Let $\A_1 \in \sat(\Spec_1)$.  By
  Theorem~\ref{thm-refinement} there exists $\A_2 \in
  \sat(\Spec_2)$ that simulates $\A_1$.  By
  Proposition~\ref{prop-TraceEquivValue}, $\prob_{\A_1}(C_w) =
  \prob_{\A_2}(C_w)$ for every word $w\in \FTr(\A_1)$.  Hence, for any
  $\omega$-regular (hence measurable) language $\lang$, one has $\prob_{\A_1}(\lang) =
  \prob_{\A_2}(\lang)$. It is in particular the case for $\V(\A_1,\obs,\fee)=\V(\A_2,\obs,\fee)$.
Therefore, $\underline{Disc}(\A_1,\obs,\fee) =
  \underline{Disc}(\A_2,\obs,\fee)$. Consequently, the theorem holds.  \qed
\end{proof}

\section{Conclusion}\label{sec:conclu}
This work investigates how refinement of probabilistic models impacts
the security --~modeled as opacity.  We provide a procedure to compute
the worst-case opacity for a subclass of IDTMCs, and we show that
opacity is monotonic with respect to simulation when implementations
are obtained through scheduling.

Directions for future work include computing disclosure for IDTMCs
with modal edges.  In addition, while we considered here only the
\emph{worst case scenario}, it would be interesting to handle both
the worst and the best case, thus providing bounds on the disclosure
of all possible implementations.

\bibliographystyle{splncs}

\begin{thebibliography}{10}

\bibitem{clarkson10}
Clarkson, M.R., Schneider, F.B.:
\newblock Hyperproperties.
\newblock Journal of Computer Security \textbf{18}(6) (September 2010)
  1157--1210

\bibitem{alur06}
Alur, R., {\v{C}}ern\'y, P., Zdancewic, S.:
\newblock Preserving secrecy under refinement.
\newblock In: Proc. of the 33rd Intl. Colloquium on Automata, Languages and
  Programming (ICALP'06). Volume 4052 of LNCS., Springer (2006)  107--118

\bibitem{mazare05}
Mazar\'e, L.:
\newblock Decidability of opacity with non-atomic keys.
\newblock In: Proc. 2nd {W}orkshop on {F}ormal {A}spects in {S}ecurity and
  {T}rust ({FAST}'04). Volume 173 of Intl. Federation for Information
  Processing., Springer (2005)  71--84

\bibitem{bryans08}
Bryans, J.W., Koutny, M., Mazar\'e, L., Ryan, P.Y.A.:
\newblock Opacity generalised to transition systems.
\newblock Intl. Jour. of Information Security \textbf{7}(6) (2008)  421--435

\bibitem{mscs15}
B{\'e}rard, B., Mullins, J., Sassolas, M.:
\newblock Quantifying opacity.
\newblock Mathematical Structures in Computer Science \textbf{25}(2) (2015)
  361--403

\bibitem{BCS15}
B{\'{e}}rard, B., Chatterjee, K., Sznajder, N.:
\newblock Probabilistic opacity for {M}arkov decision processes.
\newblock Inf. Process. Lett. \textbf{115}(1) (2015)  52--59

\bibitem{baier05}
Baier, C., Katoen, J.P., Hermanns, H., Wolf, V.:
\newblock Comparative branching-time semantics for {M}arkov chains.
\newblock Information and Computation \textbf{200} (2005)  149--214

\bibitem{larsen-jonsson}
Jonsson, B., Larsen, K.G.:
\newblock Specification and refinement of probabilistic processes.
\newblock In: LICS, IEEE Computer Society (1991)  266--277

\bibitem{segala95}
Segala, R.:
\newblock Modeling and {V}erification of {R}andomized {D}istributed
  {R}eal-{T}ime {S}ystems.
\newblock PhD thesis, MIT, Department of Electrical Engineering and Computer
  Science (1995)

\bibitem{ChatterjeeHenzinger2008}
Chatterjee, K., Henzinger, T., Sen, K.:
\newblock Model-checking omega-regular properties of interval markov chains.
\newblock In Amadio, R.M., ed.: Foundations of Software Science and Computation
  Structure (FoSSaCS) 2008. (2008)  302--317

\bibitem{Worrell2013}
Benedikt, M., Lenhardt, R., Worrell, J.:
\newblock {LTL} model checking of interval markov chains.
\newblock In: TACAS. Volume 7795 of LNCS., Springer (2013)  32--46

\bibitem{BillingsleyMeasure95}
Billingsley, P.:
\newblock {Probability and Measure}. 3rd edn.
\newblock Wiley, New York, NY (1995)

\bibitem{vardi85}
Vardi, M.Y.:
\newblock Automatic verification of probabilistic concurrent finite-state
  programs.
\newblock In: Proceedings of 26th Annual Symposium on Foundations of Computer
  Science (FOCS), IEEE Computer Society (1985)  327--338

\bibitem{courcoubetis95}
Courcoubetis, C., Yannakakis, M.:
\newblock The complexity of probabilistic verification.
\newblock {Journal of the ACM} \textbf{42}(4) (1995)  857--907

\bibitem{qest10}
B{\'e}rard, B., Mullins, J., Sassolas, M.:
\newblock Quantifying opacity.
\newblock In Ciardo, G., Segala, R., eds.: Proceedings of the 7th International
  Conference on Quantitative Evaluation of Systems (QEST'10), IEEE Computer
  Society (September 2010)  263--272

\bibitem{chaum88}
Chaum, D.:
\newblock The dining cryptographers problem: unconditional sender and recipient
  untraceability.
\newblock Journal of Cryptology \textbf{1} (1988)  65--75

\bibitem{bhargava05}
Bhargava, M., Palamidessi, C.:
\newblock Probabilistic anonymity.
\newblock In Abadi, M., de~Alfaro, L., eds.: Proceedings of the 16th
  International Conference on Concurrency Theory (CONCUR'05). Volume 3653 of
  LNCS. (2005)  171--185

\bibitem{Piterman07}
Piterman, N.:
\newblock From nondeterministic {B}{\"{u}}chi and {S}treett automata to
  deterministic parity automata.
\newblock Logical Methods in Computer Science \textbf{3}(3) (2007)

\end{thebibliography}

\newpage
\appendix

\section{Computing the probabilistic disclosure of a specification}
\label{app:computDisc}


\subsection{Detecting modal edges.}



The detection of modal edges is the first step toward computation of the disclosure of an IDTMC.

\begin{proposition}
The set of modal edges can be computed in time polynomial in the number of edges.
\end{proposition}

\begin{proof}
The decision procedure for each edge is as follows:
\begin{itemize}
\item if an edge is not weighted by an interval containing $0$, it is not modal;
\item otherwise, compute the sum of maximal interval values of all other edges stemming from the same state;
\begin{itemize}
\item if this sum is $>1$, the edge is modal;
\item if this sum is $<1$, the edge is not modal;
\item otherwise (the sum is $=1$), the edge is modal if, and only if, all intervals of other outgoing edges are closed on the right.\qed
\end{itemize}
\end{itemize}
\end{proof}

Note that the procedure does not rely entirely on the syntactic
criterion of an interval closed on $0$: it is sufficient but may lead
to false positives.  For example, consider a state with two outgoing
edges $e_1,[\frac14;\frac23]$ and $e_2,[0;1]$.  The $e_2$ edge is not
actually modal since any probability distribution satisfying the
specification can give at most $\frac23$ to $e_1$, hence must at least give weight $\frac13$ to $e_2$.  This is avoided
by the pre-computation of the least possible probability that can be
put on an edge.

\subsection{IDTMC without modal edges.}

We now prove Theorem~\ref{thm:computNoModal}.

  Note that intervals may be closed or open on any non-zero bound, which
  is not the case of IDTMCs in~\cite{ChatterjeeHenzinger2008} where all intervals are closed.  Hence
  although our procedure uses some of the ideas of Chatterjee \emph{et
    al}~\cite{ChatterjeeHenzinger2008}, it is adapted to deal with
  open intervals.

  First remark that there exists a regular language $K$ such that for
  any scheduler $A$, $\Tr(\Spec(A))=K$.  This is only true because
  $\Spec$ is assumed non-modal.  Let $\A_K$ be a PTS such that
  $\Tr(\A_K)=K$; it can be chosen of size $|\Spec|$.  By the definition of disclosure, if the secret
  $\varphi$ is $\omega$-regular and the observation function $\obs$ is
  a projection (see Section~\ref{sec:bg-opacity}), then finding the
  supremum of the disclosure means finding the maximal probability to
  reach an $\omega$-regular set of runs, namely
  $\V(\A_K,\obs,\fee)$.

  Then open intervals can be safely ignored when trying to optimize the probability of $\V(\A_K,\obs,\fee)$, \emph{i.e.} they are
  treated as closed ones.  Indeed, if the optimal scheduler uses a
  value $x$ which is the bound of an open interval, then one can build
  a family of schedulers using value $x\pm\frac1{2^n}$ for the $n$th
  scheduler.  The limit probability of reaching
  $\V(\A_K,\obs,\fee)$ is therefore the one computed when using
  exact value $x$.  Remark that using closed intervals may introduce
  intervals containing $0$, although it is of no concern since the
  observation classes are already defined and may not change, only
  their probability may change.  Said otherwise, this does not mean
  that we are computing disclosure of the closed version, since it is
  only a probability.  On the example of
  \figurename~\ref{fig:noModalImdp}, it means trying to compute the
  maximal probability of the empty set, which is indeed zero.

The procedure is hence as follows.
  Starting from a DPA $\A_\fee$ for $\fee$, a DPA $\A_\V$ for
  $\V(\A_K,\obs,\fee)$ can be built, with size exponential in the size
  of $\Spec$ and $\A_\fee$ (and with a number $k$ of colors polynomial in the
  size of $\A$ and $\A_\fee$).  This construction relies on
  intersections and complementations of DPA, with a determinization
  step that brings the exponential blowup~\cite{Piterman07}.

The construction of~\cite{ChatterjeeHenzinger2008} yields a memoryless scheduler, although it is memoryless on the product, and hence is finite-memory on the original IDTMC.
The procedure of~\cite{ChatterjeeHenzinger2008} is in EXPTIME with respect to the size of its input, hence computation of disclosure is doubly exponential: $2^{2^{O\left(|\A|\times|\A_{\varphi}|\right)}}$.

\newpage
\section{Proof of Theorem~\ref{thm-refinement}}\label{app:refinement}

  Let $\Spec_1=(S_1, s_{init,1}, T_1, \lambda_1)$ and $\Spec_2=(S_2, s_{init,2}, T_2 \lambda_2)$ be interval-based specifications such
  that $\Spec_2$ simulates $\Spec_1$ with $\Rel$. Let $sat_1$ be the
  satisfaction relation related to $A_1$ and $\A_1 = \Spec_1(A_1) = (Q_1,q_{init,1},\Delta_1,L_1)$. Then
  we show that there exists $A_2 \in Sched(\Spec_2)$ and a simulation
  relation $\Rel'$ such that $ sat_1  \circ \Rel = \Rel' \circ sat_2$ where $sat_2$ is the satisfaction
  relation related to $A_2$. 
 Then we define the scheduler $A_2$ by:
$$A_2(\rho_2)(s'_2) = \sum_{\rho_1 \in sim(\rho_2)} \mu_{\rho_2}(\rho_1) 
\sum_{s'_1 \in S_1} A_1(\rho_1)(s'_1) \cdot \delta(s'_1)(s'_2)$$ for
any $s'_2 \in S_2$, writing $\A_2 = \Spec_2(A_2) =  (Q_2,q_{init,2},\Delta_2,L_2)$.  Now the relation $\Rel'$ can be defined as
$\sim$ that relates runs that are similar ``step by step'', as defined above. To see that the conditions are satisfied, let $\rho_1$ and
$\rho_2$ be runs in $Q_1$ and $Q_2$ respectively. Then the distribution
$\delta'$ is obtained by: $$\delta'(\rho_1)(\rho_2) = 
\left\{ \begin{array}{l@{ }l}
\mu_{\rho_2}(\rho_1) \delta(\lst(\rho_1))(\lst(\rho_2)) \ &  \ 
\mbox{if} \ \rho_1 \sim \rho_2 , \\
0 & \ \mbox{otherwise.}
\end{array}\right.$$

Since $\A_1$ and $\A_2$ are PTSs, we just need to show that 
Equation (\ref{simrel-probaPTSs}) holds. 

\begin{eqnarray*}
\Delta_2(\rho_2)(\rho_2') &=& A_2(\rho_2)(s_2') \\
&=& \sum_{\rho_1 \in sim(\rho_2)} \mu_{\rho_2}(\rho_1) \sum_{s'_1 \in S_1} A_1(\rho_1)(s'_1) \cdot \delta(s'_1)(s'_2) \\
& &  \mbox{ and since } \rho'_2 = \rho_2 \cdot s_2' \\
&=&  \sum_{\rho_1 \in sim(\rho_2)}  \sum_{s'_1 \in S_1} A_1(\rho_1)(s'_1) \cdot \delta'(\rho_1 \cdot s'_1)(\rho'_2) \\
&=&  \sum_{\rho_1' \in Q_1} A_1(\rho_1)(s'_1) \cdot \delta'(\rho_1')(\rho'_2) \\
& & \mbox{by defining } \rho_1' = \rho_1 \cdot s'_1 \mbox{ and remarking that } \\
& & \delta'=0 \mbox{ if its arguments are not similar runs} \\
\Delta_2(\rho_2)(\rho_2') &=&  \sum_{\rho_1' \in Q_1} \Delta_1(\rho_1)(\rho'_1) \cdot \delta'(\rho_1')(\rho'_2). 
\hspace{4.75cm}\qed
\end{eqnarray*} 
 

\newpage
\section{Proof of Proposition~\ref{prop-TraceEquivValue}}\label{app:proof:prop-TraceEquivValue}
Assume by induction that the proposition holds for
  every word of length $n$.  Let $w \in \FTr(\A_1) = \FTr(\A_2)$ (recall Proposition~\ref{prop-TraceEquiv}) of length $n+1$ with $w = w_0 a$  for some $a \in \Sigma$.  A run $\rho'$ of
  $\A_2$ that produces $w$ can be assumed to be of the form  $\rho' = \rho_0' s_2'$ with $\tr(\rho_0') = w_0$ and  $\lambda(s_2') = a$.
	Then $\prob_{\A_2}(C_{\rho'}) = \prob_{\A_2}(C_{\rho_0'}) \Delta_2(s_2)(s_2')$ where  $s_2 = \lst(\rho_0')$
  and hence 
  \begin{eqnarray*}
\prob_{\A_2}(C_w) &=& \sum_{\substack{\rho' \in \FRuns(\A_2)\\\tr(\rho')=w}} \prob_{\A_2}(C_{\rho'}) \\
                  &=& \sum_{\substack{s_2,s_2'\in S_2}}\hspace*{-1em}\sum_{\substack{\rho_0' \in \FRuns(\A_2)\\\tr(\rho_0')=w_0,\\ \lst(\rho_0')=s_2, \lst(\rho')=s_2'}}\hspace*{-2em} \prob_{\A_2}(C_{\rho_0'}) \Delta_2(s_2)(s_2') \\
\end{eqnarray*}
 Now let $\A_1$ s.t.  $\A_2$ simulates  $\A_1$ then, as 
    \[\sum_{\substack{\rho_0 \in \FRuns(\A_1)\\\rho_0 \sim \rho_0'\\\lst(\rho_0)=s_1}} \mu_{\rho_0'}(\rho_0) =  \sum_{s_1 \in S_1} \sum_{\substack{\rho_0 \in \FRuns(\A_1)\\\rho_0 \sim \rho_0'\\\lst(\rho_0)=s_1}}\hspace*{-1em} \mu_{\rho_0'}(\rho_0) = 1\]
   we get that 
\begin{eqnarray*}
\prob_{\A_2}(C_w)  &=& \sum_{\substack{s_2,s_2'\in S_2}} \hspace*{-1em}\sum_{\substack{\rho_0' \in \FRuns(\A_2)\\\tr(\rho_0')=w_0,\\ \lst(\rho_0')=s_2, \lst(\rho')=s_2'}}\hspace*{-2em} \prob_{\A_2}(C_{\rho_0'}) \Delta_2(s_2)(s_2') \cdot \\&&\hspace{4.5cm}\sum_{s_1 \in S_1} \sum_{\substack{\rho_0 \in \FRuns(\A_1)\\\rho_0 \sim \rho_0'\\\lst(\rho_0)=s_1}}\hspace*{-1em} \mu_{\rho_0'}(\rho_0) \\
\end{eqnarray*}

\begin{eqnarray*}
\prob_{\A_2}(C_w) &=& \sum_{\substack{s_1 \in S_1\\s_2,s_2'\in S_2}}\sum_{\substack{\rho_0' \in \FRuns(\A_2)\\\tr(\rho_0')=w_0,\\ \lst(\rho_0')=s_2, \lst(\rho')=s_2'}}\hspace*{-2em} \prob_{\A_2}(C_{\rho_0'}) \Delta_2(s_2)(s_2')  \sum_{\substack{\rho_0 \in \FRuns(\A_1)\\\rho_0 \sim \rho_0'\\\lst(\rho_0)=s_1}}\hspace*{-1em} \mu_{\rho_0'}(\rho_0) \\
& & \mbox{As   the terms are null if it is not the case that  $s_1 \Rel s_2$,  we have:}  \\
                  &=&  \sum_{\substack{s_1 \in S_1\\s_2,s_2'\in S_2}}\hspace*{-0.66em}\sum_{\substack{\rho_0' \in \FRuns(\A_2)\\\tr(\rho_0')=w_0,\\ \lst(\rho_0')=s_2, \lst(\rho')=s_2'}}\hspace*{-2em} \prob_{\A_2}(C_{\rho_0'}) \sum_{s'_1 \in S_1} \Delta_1(s_1)(s'_1)\cdot \\&&\hspace*{6.5cm}\delta_{s_1,s_2} (s'_1)(s'_2)  \hspace*{-0.5em}\sum_{\substack{\rho_0 \in \FRuns(\A_1)\\\rho_0 \sim \rho_0'\\\lst(\rho_0)=s_1}}\hspace*{-1em} \mu_{\rho_0'}(\rho_0) \\
& & \mbox{And as  the terms are null if $\lambda(s_1') \neq \lambda(s_2') = \lambda(\lst(\rho))$, we have:}  \\
                  &=&  \sum_{\substack{s_1 \in S_1\\s_2,s_2'\in S_2}}\hspace*{-0.66em}\sum_{\substack{\rho_0' \in \FRuns(\A_2)\\\tr(\rho_0')=w_0,\\ \lst(\rho_0')=s_2, \lst(\rho')=s_2'}}\hspace*{-2em} \prob_{\A_2}(C_{\rho_0'}) \hspace*{-1em}\sum_{\substack{s'_1 \in S_1\\\lambda(s_1) = \lambda(\lst(\rho))}}\hspace*{-1.5em} \Delta_1(s_1)(s'_1)\cdot \\&&\hspace{6.66cm} \delta_{s_1,s_2} (s'_1)(s'_2)  \hspace*{-1em}\sum_{\substack{\rho_0 \in \FRuns(\A_1)\\\rho_0 \sim \rho_0'\\\lst(\rho_0)=s_1}}\hspace*{-1.2em} \mu_{\rho_0'}(\rho_0) \\
                  &=& \sum_{s_1 \in S_1} \sum_{\substack{s'_1 \in S_1\\\lambda(s_1) = \lambda(\lst(\rho))}} \sum_{\substack{s_2\in S_2\\s_2'\in S_2}} \delta_{s_1,s_2} (s'_1)(s'_2) \cdot \\&&\hspace{2.75cm}\left(\sum_{\substack{\rho_0' \in \FRuns(\A_2)\\\tr(\rho_0')=w_0,\\ \lst(\rho_0')=s_2, \lst(\rho')=s_2'}}\hspace*{-2em} \prob_{\A_2}(C_{\rho_0'}) \Delta_1(s_1)(s'_1) \sum_{\substack{\rho_0 \in \FRuns(\A_1)\\\rho_0 \sim \rho_0'\\\lst(\rho_0)=s_1}} \hspace*{-1em}\mu_{\rho_0'}(\rho_0)\right) \\
                                    & & \mbox{And since $\sum_{s_2'\in S_2} \delta_{s_1,s_2} (s'_1)(s'_2) = 1$, we have:} \\
                  &=& \sum_{s_1 \in S_1} \hspace*{-1em}\sum_{\substack{s'_1 \in S_1\\\lambda(s_1) = \lambda(\lst(\rho))}}\hspace*{-1em} \sum_{s_2\in S_2} \sum_{\substack{\rho_0' \in \FRuns(\A_2)\\\tr(\rho_0')=w_0,\\ \lst(\rho_0')=s_2}} \hspace*{-1em}\prob_{\A_2}(C_{\rho_0'}) \Delta_1(s_1)(s'_1) \hspace*{-1em}\sum_{\substack{\rho_0 \in \FRuns(\A_1)\\\rho_0 \sim \rho_0'\\\lst(\rho_0)=s_1}}\hspace*{-1em} \mu_{\rho_0'}(\rho_0) \\
\prob_{\A_2}(C_w) &=& \sum_{s_1 \in S_1} \hspace*{-1em}\sum_{\substack{s'_1 \in S_1\\\lambda(s_1) = \lambda(\lst(\rho))}}\hspace*{-0.25em} \sum_{\substack{\rho_0' \in \FRuns(\A_2)\\\tr(\rho_0')=w_0}} \hspace*{-1em}\prob_{\A_2}(C_{\rho_0'}) \Delta_1(s_1)(s'_1) \hspace*{-1em}\sum_{\substack{\rho_0 \in \FRuns(\A_1)\\\rho_0 \sim \rho_0'\\\lst(\rho_0)=s_1}}\hspace*{-1em} \mu_{\rho_0'}(\rho_0) \\
\end{eqnarray*}

\begin{eqnarray*}
\prob_{\A_2}(C_w) &=& \sum_{s_1 \in S_1} \hspace*{-0.75em}\sum_{\substack{s'_1 \in S_1\\\lambda(s_1) = \lambda(\lst(\rho))}} \hspace*{-1em} \Delta_1(s_1)(s'_1) \hspace*{-0.75em}\sum_{\substack{\rho_0' \in \FRuns(\A_2)\\\tr(\rho_0')=w_0}} \sum_{\substack{\rho_0 \in \FRuns(\A_1)\\\rho_0 \sim \rho_0'\\\lst(\rho_0)=s_1}}\hspace*{-0.75em} \prob_{\A_2}(C_{\rho_0'}) \mu_{\rho_0'}(\rho_0) \\
                  &=& \sum_{s_1 \in S_1} \hspace*{-0.75em}\sum_{\substack{s'_1 \in S_1\\\lambda(s_1) = \lambda(\lst(\rho))}} \hspace*{-1em} \Delta_1(s_1)(s'_1) \hspace*{-0.75em}\sum_{\substack{\rho_0' \in \FRuns(\A_2)\\\tr(\rho_0')=w_0}} \\&&\hspace{4.5cm}\sum_{\substack{\rho_0 \in \FRuns(\A_1)\\\rho_0 \sim \rho_0'\\\lst(\rho_0)=s_1}}\hspace*{-0.75em} \prob_{\A_2}(C_{\rho_0'}) \frac{\prob_{\A_1}(C_{\rho_0})}{\prob_{\A_1}(sim(\rho_0'))} \\
                  &=& \sum_{s_1 \in S_1} \hspace*{-0.75em}\sum_{\substack{s'_1 \in S_1\\\lambda(s_1) = \lambda(\lst(\rho))}} \hspace*{-1em} \Delta_1(s_1)(s'_1) \hspace*{-0.75em}\sum_{\substack{\rho_0 \in \FRuns(\A_1)\\\tr(\rho_0)=w_0\\\lst(\rho_0)=s_1}} \frac{\prob_{\A_1}(C_{\rho_0})}{\prob_{\A_1}(sim(\rho_0'))} \cdot \\&&\hspace*{7cm} \sum_{\substack{\rho_0' \in \FRuns(\A_2)\\\tr(\rho_0')=w_0}} \hspace*{-1em} \prob_{\A_2}(C_{\rho_0'}) \\
                  &=& \sum_{s_1 \in S_1} \hspace{-0.75em}\sum_{\substack{s'_1 \in S_1\\\lambda(s_1) = \lambda(\lst(\rho))}} \hspace{-1em} \Delta_1(s_1)(s'_1) \hspace{-0.75em}\sum_{\substack{\rho_0 \in \FRuns(\A_1)\\\tr(\rho_0)=w_0\\\lst(\rho_0)=s_1}} \frac{\prob_{\A_1}(C_{\rho_0})}{\prob_{\A_1}(sim(\rho_0'))} \cdot \prob_{\A_2}(C_{w_0}) \\
                  &&\mbox{and by induction hypothesis, } \prob_{\A_2}(C_{w_0}) = \prob_{\A_1}(C_{w_0}) \mbox{:}\\
                  &=& \sum_{s_1 \in S_1} \hspace{-0.75em}\sum_{\substack{s'_1 \in S_1\\\lambda(s_1) = \lambda(\lst(\rho))}} \hspace{-1em} \Delta_1(s_1)(s'_1) \hspace{-0.75em}\sum_{\substack{\rho_0 \in \FRuns(\A_1)\\\tr(\rho_0)=w_0\\\lst(\rho_0)=s_1}} \frac{\prob_{\A_1}(C_{\rho_0})}{\prob_{\A_1}(sim(\rho_0'))} \cdot \prob_{\A_1}(C_{w_0}) \\
                  &&\mbox{and since } \prob_{\A_1}(sim(\rho_0')) = \prob_{\A_1}(C_{w_0}) \mbox{:}\\
                  &=& \sum_{s_1 \in S_1} \hspace{-0.75em}\sum_{\substack{s'_1 \in S_1\\\lambda(s_1) = \lambda(\lst(\rho))}} \hspace{-0.75em} \Delta_1(s_1)(s'_1) \hspace{-0.75em}\sum_{\substack{\rho_0 \in \FRuns(\A_1)\\\tr(\rho_0)=w_0\\\lst(\rho_0)=s_1}} \hspace{-0.75em}\prob_{\A_1}(C_{\rho_0}) \\
                  &=& \sum_{s_1 \in S_1} \sum_{\substack{\rho_0 \in \FRuns(\A_1)\\\tr(\rho_0)=w_0\\\lst(\rho_0)=s_1}} \hspace{-0.75em}\prob_{\A_1}(C_{\rho_0}) \hspace{-0.75em}\sum_{\substack{s'_1 \in S_1\\\lambda(s_1) = \lambda(\lst(\rho))}}\hspace{-1.5em}  \Delta_1(s_1)(s'_1)\\
                  &=& \sum_{\substack{\rho_0 \in \FRuns(\A_1)\\\tr(\rho_0)=w_0}}\hspace{-0.75em} \prob_{\A_1}(C_{\rho_0}) \Delta_1(\lst(\rho_0))(\lst(\rho))\\
\prob_{\A_2}(C_w) &=& \prob_{\A_1}(C_w)\hspace{9cm}\qed
\end{eqnarray*}

\end{document}